\documentclass[12pt]{article}
\begin{document}
\begin{titlepage}
\begin{flushright}
KEK-TH-1005
\end{flushright}
\begin{center}
{\Large\bf Structure of Split Supersymmetry and Simple Models}
\end{center}
\vspace{1cm}
\begin{center}
Naoyuki {Haba}$^{(a),}$ 
\footnote{E-mail: haba@ias.tokushima-u.ac.jp}
and 
Nobuchika {Okada}$^{(b),}$
\footnote{E-mail: okadan@post.kek.jp},
\end{center}
\vspace{0.2cm}
\begin{center}
${}^{(a)}$ {\it Institute of Theoretical Physics, University of
Tokushima, Tokushima 770-8502, Japan}
\\[0.2cm]
${}^{(b)}$ {\it Theory Division, KEK, Tsukuba 305-0801, 
Japan}
\end{center}
\vspace{1cm}
\begin{abstract}

We derive in detail a condition on the Higgs mass parameters 
 that is necessary for the recently proposed ``split supersymmetry'' 
 (split SUSY) scenario to provide a realistic magnitude of $\tan \beta$. 
The nature of this condition can be understood by showing 
 how the Higgs sector of the minimal supersymmetric Standard Model 
 reduces to that of the Standard Model 
 in the heavy limit of the soft supersymmetry breaking 
 Higgs mass parameters. 
Based on this condition, 
 we present some simple supersymmetry breaking models 
 that each provides a realistic split-SUSY mass spectrum,
 in accordance with the scale of the gravitino mass ($m_{3/2}$) 
 in relation to those of the soft scalar mass ($\tilde{m}$) 
 and the gaugino mass ($M_{1/2}$) employed in each, 
 namely $m_{3/2} \geq \tilde{m}$, $\tilde{m} \geq m_{3/2} \geq M_{1/2}$ 
 and $M_{1/2} \geq m_{3/2}$, respectively, 
 with the relation $\tilde{m} \gg M_{1/2}$ of the split-SUSY mass spectrum. 

\end{abstract}
\end{titlepage}
\newpage

\section{Introduction}

The Standard Model (SM),
 with a simple extension to incorporate neutrino masses and mixings, 
 is in good agreement with 
 almost of all current experimental data.
However, the SM contains the gauge hierarchy problem 
 of quantum field theory. 
This results from the quadratic divergence of the 
 Higgs boson mass on the new physics scale 
 arising in quantum theories,  
 which makes a very precise fine-tuning 
 necessary in in order to realize the correct electroweak scale 
 if this new physics scale lies at a high energy scale, 
 such as the scale of the grand unified theory (GUT) 
 or the Planck scale. 
In other words, the vacuum in the SM 
 is not stable with respect to quantum corrections. 
It is well known that this fine-tuning problem can be solved 
 by introducing supersymmetry (SUSY)~\cite{SUSY-review}. 
The minimally extended SUSY SM (MSSM) has the elegant feature 
 of gauge coupling unification, and for this reason, 
 many people believe that there exists a 4-dimensional SUSY. 
Some people also believe that SUSY is required 
 for the construction of a quantum theory of gravity. 

However, SUSY particle has not yet been observed experimentally. 
Also, the proton-decay predicted by SUSY GUT models 
 has not yet been observed~\cite{Eidelman:2004wy}. 
Given this situation, we might consider the possibility
 of heavy SUSY particles as one possibility. 
Recently, the split supersymmetry (split SUSY) scenario
 was proposed~\cite{Split-SUSY, Split-SUSY2}. 
In this scenario, nature is fine-tuned intrinsically, 
 and SUSY has nothing to do with the gauge hierarchy problem. 
The scalar masses are super heavy, 
 while the fermion masses are maintained 
 at the electroweak scale, 
 protected by the chiral ($U(1)_R$) symmetry.
In this way, 
 the split SUSY scenario forgets the fine-tuning problem 
 originating in the Higgs mass quadratic divergence. 
Related studies are given in Ref.~\cite{Kokorelis:2004dc}.

In this paper, we first overview the fact
 that the Higgs sector of the MSSM 
 reduces to that of the SM 
 in the heavy limit of the soft SUSY breaking 
 Higgs mass parameters.  
This means that split SUSY is just
 the MSSM containing super-heavy scalar masses. 
We examine the Higgs potential in detail and 
 derive a condition on the Higgs mass parameters 
 ($|{m}_{u}^2|\sim |{m}_{d}^2|\sim |B\mu |$) 
 that is necessary for the split-SUSY scenario 
 to yield suitable values of $\tan \beta$, 
 i.e., those in the range $1 \leq \tan\beta \leq 60$. 
Then, we present some simple supersymmetry breaking models 
 that provide realistic split-SUSY mass spectra, 
 in accordance with the relations among the scales 
 of the gravitino mass ($m_{3/2}$), 
 the soft scalar mass ($\tilde{m}$) 
 and the gaugino mass ($M_{1/2}$) used in each, 
 namely 
 $m_{3/2} \geq \tilde{m}$, $\tilde{m} \geq m_{3/2} \geq M_{1/2}$ 
 and $M_{1/2} \geq m_{3/2}$, respectively, 
 along with the condition $\tilde{m} \gg M_{1/2}$ 
 of the split-SUSY mass spectrum.

\section{Structure of split-SUSY}

Let us present the detailed structure of 
 the Higgs potential in split SUSY.
This is the same as the minimal SUSY standard model
 (MSSM) with super-heavy soft masses.
Here we explicitly give the fine-tuning conditions 
 required in the Higgs potential of split SUSY. 

The standard model (SM) Higgs potential is
 given by
\begin{equation}
{V_{\rm SM}}=-m^2 h^\dagger h +{\lambda \over 2} (h^\dagger h)^2,
\label{SM}
\end{equation}
while that of 
 the MSSM is given by
\begin{eqnarray}
{V_{\rm MSSM}}&=& m_u^2|H_u|^2+
 m_d^2|H_d|^2+
 (B \mu \epsilon_{ij}H_u^iH_d^j+ \mbox{h.c.}) \nonumber \\
&& +{g^2+g'^2 \over8}(|H_u|^2-|H_d|^2)+
   {g^2 \over2}|H_u|^2|H_d|^2-
   {g^2 \over2}|\epsilon_{ij}H_u^iH_d^j|^2 ,
\label{MSSM}
\end{eqnarray}
where $m_u^2 = |\mu|^2+\tilde{m}_u^2$ and
 $m_d^2 = |\mu|^2+\tilde{m}_d^2$. 
The masses $\tilde{m}_{u,d} \simeq \tilde{m} $ are 
 the soft SUSY breaking masses 
 of the up-type and down-type Higgs doubles.
We make $B \mu$ real and positive through field redefinitions.
Each neutral component of the Higgs doublets 
 develops a vacuum expectation value (VEVs),
 $\langle H_u \rangle =v \sin \beta / \sqrt{2}$ and
 $\langle H_d \rangle =v \cos \beta / \sqrt{2}$. 
In the split-SUSY scenario, 
 the threshold corrections to the quartic 
 coupling are small, 
 due to the smallness of the $A$ terms, and 
 the energy scale dependence of the quartic coupling 
 should be estimated using the renormalization 
 group equation analysis
 rather than the effective
 potential~\cite{Split-SUSY}. 
Thus, obtaining the tree-level 
 Higgs potential is
 enough for the analysis 
 at the high energy scale of $\tilde{m}$. 
As shown below, 
 because of the large soft SUSY breaking terms, 
 only the SM-like Higgs scalar survives at low energies, 
 and the MSSM Higgs potential is reduced to the SM one. 

Necessary and sufficient conditions 
 for realizing a suitable electroweak symmetry breaking 
 are given by~\cite{Split-SUSY} 
\begin{eqnarray}
\label{5a}
&& m_u^2 + m_d^2 -2 B\mu >0, \;\;\;\;\;\;\;
    \left({m_u^2 + m_d^2}\right)^2 < 
    \left({m_u^2 - m_d^2}\right)^2 + (2\mu B)^2, \nonumber \\
&&  \;\;\;\;\;\;\;\;\;\;\;\;\;\;\;\;\;\;\;\;\;\;\;\;\;\;\;\;
\left({m_u^2 + m_d^2}+ m_{EW}^2 \right)^2 > 
    \left({m_u^2 - m_d^2}\right)^2 + (2\mu B)^2, 
\end{eqnarray}
where $m_{EW}$ is the electroweak mass scale, 
 which is ${\mathcal O}(10^2)$ GeV. 
The third condition implies that the magnitude 
 of the negative mass squared eigenvalue 
 should be the electroweak scale. 
It is well known that 
 the minimization conditions $dV/dH_{u,d}=0$ 
 can be expressed as 
\begin{eqnarray}
\label{B}
&& \sin 2\beta ={2 B\mu \over m_u^2 + m_d^2}, \;\;\;
 M_Z^2={m_u^2 - m_d^2 \over \cos 2 \beta}-(m_u^2 + m_d^2).
\end{eqnarray}
We now introduce a field redefinition, 
 employing $H_1\equiv(\epsilon H_d^*)$ and $H_2\equiv H_u$. 
Then, in the basis $(H_1,H_2)$, 
 the Higgs mass matrix is obtained as 
\begin{eqnarray}
\left(
\begin{array}{cc}
m_d^2 & B \mu \\
B \mu & m_u^2 
\end{array}
\right) = \left(
\begin{array}{cc}
m_A^2 \sin^2\beta -{1\over2}M_Z^2\cos(2\beta) & m_A^2\sin\beta\cos\beta \\
m_A^2\sin\beta\cos\beta & m_A^2\cos^2 \beta+{1\over2}M_Z^2\cos(2\beta)
\end{array}
\right) , 
\label{HM2}
\end{eqnarray} 
by using Eq.~(\ref{B}). 
Here we have $m_A^2 \equiv m_u^2 + m_d^2$.
This mass matrix automatically satisfies the conditions in Eq.~(\ref{5a}) 
 with the identification  $m_{EW}=M_Z$. 

In the split-SUSY scenario, 
 one linear combination
 of the Higgs
 doublet, 
\begin{equation}
 \widetilde{h}=-\cos\beta {H_1} + \sin \beta H_2,
\label{H}
\end{equation}
is light and only survives 
 below the energy scale of 
 ${\mathcal O}(\tilde{m})$,
 as shown in Appendix A.
(The eigenstate orthogonal to $\widetilde{h}$ is 
 $\widetilde{H}=-\sin\beta {H_1} -\cos \beta H_2$\footnote{
In the basis $(\widetilde{h},\widetilde{H})$,
 Eq.~(\ref{HM2}) is rewritten as  
$
\left(
\begin{array}{cc}
-{1\over2}M_Z^2 \cos^2(2\beta) & -{M_Z^2\over 4}\sin(4\beta) \\
-{M_Z^2\over 4}\sin(4\beta) & 
 m_A^2 + {M_Z^2\over4}(1+\cos(4\beta)) 
\end{array}
\right). 
$
}.)
For this reason, the low energy effective theory
 should be written in terms of $\widetilde{h}$ only.
The effective Higgs potential 
 is obtained from Eqs.~(\ref{MSSM}) and (\ref{H}) as
\begin{eqnarray}
{V^{\rm eff}_{\rm MSSM}}&=& -m'^2|\widetilde{h}|^2+
 {\lambda'\over2}|\widetilde{h}|^4,\;\;\;\;\;
 \left(\lambda'={g^2+g'^2 \over 4}\cos^2 2 \beta \right),
\label{MSSM3}
\end{eqnarray}
where $m'^2\equiv M_Z^2\cos^2(2\beta)$ [see Eq.~(\ref{exact})].
The Higgs mass becomes zero (resp., $M_Z$)
 when $\tan \beta=1$ (resp., $\beta=\pi/2$) 
 at the high energy scale of ${\mathcal O}(\tilde{m})$.
This can also be understood by considering 
 the effective quartic coupling, $\lambda'$, in Eq.~(\ref{MSSM3})
 as follows.
When $\tan\beta=1$, 
 $\lambda'$ is zero, and therefore 
 the Higgs mass $\sqrt{\lambda} v$ vanishes
 at the SUSY breaking scale, and only the 
 radiative corrections induce a finite
 Higgs mass at the low energy. 
Contrastingly, when $\beta=\pi/2$, 
 the Higgs mass becomes $M_Z$ at the SUSY breaking scale, 
 and in this case also, the mass is increased by the radiative corrections.
This is the reason why $\cos 2\beta=0$ (resp., $\cos2\beta=-1$)
 is found to have the smallest (resp., largest) mass of the 
 low energy physical Higgs scalar, $h^0$,
 in Ref.~\cite{Split-SUSY}. 

We have shown that the MSSM Higgs potential
 reduces to the SM one 
 when the soft scalar masses, $\tilde{m}$,
 are much larger than the electroweak scale. 
Explicitly, 
 the Higgs doublet, $\widetilde{h}$, is the direction 
 of the VEV and also contains all would-be NG bosons and
 one SM-like physical Higgs scalar. 
The important point is that 
 the vacuum stability conditions in Eq.~(\ref{B}) 
 must be satisfied 
 even when we introduce super-heavy soft masses.
It should be noted that the conditions given in Eq.~(\ref{B}) 
 are essential and that these represent 
 the fine-tuning required in the split-SUSY scenario. 
Some examples of the split-SUSY scenario
 given in Ref.~\cite{Split-SUSY} suggest 
 scalar masses of ${\mathcal O}(10^{12-13})$ GeV
 (which is the scale favored by the cosmological considerations), 
 while $B\mu$ is suppressed by the chiral [$U(1)_R$] symmetry.
However, in this case, there is an extremely large 
 $\tan\beta \sim m_d^2/B\mu$, 
 as seen from 
 Eqs.~(\ref{B}) and (\ref{HM2}), 
 and therefore it is difficult to obtain 
 a realistic bottom quark Yukawa coupling. \footnote{
 There is a finite quantum correction for the 
 bottom quark mass, which is produced through 
 the anti-holomorphic Yukawa interaction induced by  
 the gluino and higgsino 1-loop diagrams~\cite{Hall:1993gn}.
 However, this correction is negligibly small, 
 due to the super-heavy 
 masses of the sfermions in the split-SUSY scenario.
}
In order to obtain a realistic value of $\tan\beta$, 
 $B\mu$ should be of the same order as the Higgs mass. 
It is non-trivial to construct a model 
 that can naturally provide a realistic split-SUSY mass spectrum.

\section{Simple models}

In this section we present simple SUSY breaking models 
 satisfying the condition 
 $|{m}_{u}^2|\sim |{m}_{d}^2|\sim |B\mu |$, 
 which is necessary for a realistic split-SUSY scenario 
 with $1 \leq \tan\beta \leq 60$, 
 as shown in the previous section. 
There are many possible ways to construct such models. 
We consider several models 
 characterized by the scale of 
 the gravitino mass, $m_{3/2}$, 
 in comparison with the soft scalar mass, $\tilde{m}$ 
 and the gaugino mass $M_{1/2}$, 
 with the split-SUSY mass spectrum, 
 $\tilde{m} \gg M_{1/2}$. \\

{\bf 1. Case of large gravitino mass ($0.01 \; m_{3/2} > M_{1/2}$)}\\

\noindent 
In this case, we should first note 
 the gaugino mass generated through the superconformal anomaly 
 (anomaly mediation)~\cite{AMSB1, AMSB2}, 
 which is approximated as 
 $(\alpha_{SM}/4 \pi) F_\phi \simeq 0.01 F_\phi $, 
 where $\alpha_{SM}$ is the gauge coupling in the Standard Model, 
 and $F_\phi$ is the F-term of the compensating multiplet. 
In normal SUSY breaking scenarios in SUGRA, 
 we obtain $F_\phi \simeq m_{3/2}$, 
 where $m_{3/2}$ is the gravitino mass. 
Therefore, for a SUSY breaking model with gravitino mass 
 satisfying $0.01 m_{3/2} > M_{1/2}$, 
 a mechanism that can suppress the anomaly mediation 
 is necessary to realize the split-SUSY scenario. 
Such a model is the ``almost no-scale'' 
 SUGRA model~\cite{Luty:2002ff, Luty:2002hj}, 
 whose structure is in fact crucial 
 in the split-SUSY scenario, 
 as shown in the original paper~\cite{Split-SUSY}. 

Let us first introduce the almost no-scale model. 
Although Refs.~\cite{Luty:2002ff, Luty:2002hj} consider 
 extra-dimensional theories, 
 a 4D SUGRA model can yield the same type of a structure 
 as these extra-dimensional theories 
 if we allow fine-tuning of the parameters in the Kahler potential. 
We consider the SUGRA Lagrangian, 
\begin{eqnarray}
{\mathcal L} = \int d^4\theta {\mathcal K}(z^\dagger ,z) 
 \phi^\dagger \phi +
\left\{ \int d^2\theta \phi^3 W_0 + {\mbox h.c.} \right\}, 
\end{eqnarray}
with the Kahler potential 
\begin{eqnarray}
  {\mathcal K}(z^\dagger, z) 
 = -3 M_4^2 (z+z^\dagger + \epsilon f(z,z^\dagger)), 
\end{eqnarray}
where $z$ and $\phi$ denote a hidden sector (dilaton) superfield 
 and a compensating multiplet ($\phi=1+F_\phi$), respectively, 
 $\epsilon$ is a small dimensionless parameter, 
 $M_4$ is the 4D Planck scale, 
 and $W_0$ is a constant superpotential. 
The original (4D) no-scale model~\cite{no-scale} 
 is obtained in the limit $\epsilon \rightarrow 0$. 
The equations of motion for the auxiliary fields, 
 $d{\mathcal L}/dF_\phi^\dagger =0$ 
 and $d{\mathcal L}/dF_z^\dagger =0$, lead to 
\begin{eqnarray}
\label{Fphi}
  F_\phi \simeq 
  -{W_0^\dagger \over M_4^2} \epsilon f_{z^\dagger z} 
  = - \epsilon m_{3/2} 
  f_{z^\dagger z}, \;\;\;\;\;
  F_z \simeq {W_0^\dagger \over M_4^2} = m_{3/2} ,  
\end{eqnarray}
 for small $\epsilon$. 
Here, $f_{z^\dagger z}$ stands for 
 $\partial^2 f(z^\dagger ,z)/ \partial z^\dagger \partial z$. 
Then, the scalar potential is given by 
\begin{eqnarray}
 V = -3 F_\phi W_0 
  \simeq 3  {|W_0|^2 \over M_4^2} \epsilon f_{z^\dagger z} 
  = 3 \epsilon m_{3/2}^2 M_4^2 f_{z^\dagger z} 
\end{eqnarray}
Assuming that $f_{z^\dagger z}$ has the form 
 $f_{z^\dagger z}= (|z|^2-1/4)^2 -1$, for example,
 the potential has a minimum at $\langle z \rangle =1/2$, 
 with potential energy 
\begin{eqnarray}
 V_{\mbox{min}} \simeq  - 3 \epsilon m_{3/2}^2 M_4^2.  
\label{no scale potential energy} 
\end{eqnarray} 
Here, the almost no-scale structure is realized; 
 that is, we $F_\phi \simeq \epsilon m_{3/2} \ll m_{3/2}$. 

Of course, the contact terms among 
 the gauginos in the visible sector and $z$ 
 should be suppressed in order to realize the split-SUSY scenario. 
However, in this model, it is difficult to find a symmetry 
 that would forbid such contact terms. 
A simple way to avoid this problem is to introduce 
 the sequestering scenario~\cite{AMSB1}, 
 in which we assume that the dilaton sector 
 and the visible sector exist on different branes that 
 are spatially separated in the extra-dimensions.~\footnote{
In such a scenario, one of the most important points 
 is radius stabilization, 
 since it is, in general, very closely related 
 to SUSY breaking and its mediation mechanism. 
With regard to this point, the models proposed in \cite{Maru-Okada} 
 are noteworthy, because in them, radius stabilization is realized 
 independently of the SUSY breaking and its mediation mechanism.} 
Because the contribution from the anomaly mediation is sub-dominant, 
 an additional SUSY breaking source and a SUSY breaking mediation mechanism 
 must be introduced in order to realize a split-SUSY mass spectrum. 
For this purpose, consider a hidden sector 
 with a $U(1)$ gauge symmetry and the Fayet-Iliopoulos D-term 
 with particles $X$ and $Y$, 
 which have $U(1)$ charges $1$ and $-1$, respectively. 
Suppose that these hidden sector fields exist 
 on the visible sector brane and 
 that there exists a superpotential $W=m X Y$. 
Together with the dilaton sector, the total Lagrangian 
 (in the 4D effective theory) is given by 
\begin{eqnarray} 
{\cal L}&=& \int d^4 \theta \left[
 {\cal K}(z^\dagger, z) \phi^\dagger \phi 
 + X^\dagger e^{+2 g V} X + Y^\dagger e^{- 2 g V} Y 
  \right]  \nonumber \\
 &+& \left[ 
  \int d^2 \theta \left( \phi^3 W_0 + \phi m X Y \right)
 +{\mbox h.c.} \right]  \nonumber \\ 
&+&\left[ \frac{1}{4} \int d^2 \theta 
 {\cal W}^\alpha {\cal W}_\alpha + {\mbox h.c.} 
 \right] 
+ \int d^4 \theta \xi^2 V, 
\end{eqnarray} 
where the last term is the Fayet-Iliopoulos D-term, 
 and $\xi$ is a real parameter with dimension of mass 
 ($\xi^2 > 0$ with our definition of the $U(1)$ charge). 
If we consider an anomalous $U(1)$ gauge theory, 
 the parameter can be understood as 
 $\xi^2 = g_S^2 M_S^2 Tr Q/192\pi^2$ 
 with the string coupling $g_S$, the string scale $M_S$, 
 and the anomalous $U(1)$ charge $Q$~\cite{AU1}. 
Here, the superfields $X$ and $Y$ have been rescaled as  
 $X, Y \rightarrow X/\phi, Y/\phi$, 
 so that the compensating multiplet $\phi$ disappears 
 in the Kahler potential for $X$ and $Y$. 

Note that the dilaton sector (the almost no-scale sector) and 
 the $U(1)$ gauge sector are decoupled in the Kahler potential. 
Because of this fact, in the equations of motion 
 for the auxiliary fields in the almost no-scale sector, 
 $W_0$ is simply replaced by $W_0 +1/3 m X Y$. 
Thus, if $ |W_0| \gg \langle | m X Y | \rangle $, 
 the structure of the almost no-scale sector remains almost the same. 
It must be noted, though, that 
 non-zero $F_\phi$ induced in the almost no-scale sector 
 affects the scalar potential for $X$ and $Y$. 
However, if $F_\phi \simeq \epsilon m_{3/2} \ll m$, 
 the scalar potential of the $U(1)$ gauge sector in SUGRA 
 is almost the same as that 
 in the global SUSY limit ($F_\phi \rightarrow 0$), 
 because the scalar potential for $X$ and $Y$ is controlled 
 by the scale $m$. 
This is the case that we examine in the following. 

Analyzing the potential in the $U(1)$ gauge sector 
 (ignoring $F_\phi$), we find 
\begin{eqnarray} 
 \langle X \rangle =0, \; \; \;  
 \langle Y \rangle = \pm \sqrt{\xi^2-\frac{m^2}{g^2}} 
\end{eqnarray} 
and 
\begin{eqnarray} 
 \langle F_X \rangle =m \langle Y \rangle, \; \; \;  
 \langle F_Y \rangle = 0 , \; \; \;  
 \langle D \rangle = \frac{m^2}{g} , 
\end{eqnarray} 
 for $\xi^2 > m^2/g^2 $. 
In the following, we assume $\xi^2 \gg m^2/g^2 $ for simplicity. 
Then the potential energy is found to be 
\begin{eqnarray} 
 V_{\mbox{min}} = m^2 \langle Y \rangle^2 +\frac{1}{2} \langle D \rangle^2 
     \simeq  m^2 \xi^2 > 0 . 
\end{eqnarray} 
In order to obtain a vanishing cosmological constant, 
 this potential energy should be canceled out 
 by the negative contribution 
 in the almost no-scale sector, 
 given by Eq.~(\ref{no scale potential energy}), 
 and hence we have 
\begin{eqnarray}  
  m^2 \xi^2 \simeq \epsilon m_{3/2}^2 M_4^2.  
 \label{vanishing cc}
\end{eqnarray}  

We next consider the soft SUSY breaking mass spectrum. 
We impose R-parity with the usual assignments for the MSSM fields 
 and even for the other fields. 
The values of the scalar soft mass squared 
 for the MSSM particles, represented by $\Psi$, 
 are determined by 
\begin{equation}
\left[{X^\dagger X \over M_4^2} 
 \Psi^\dagger \Psi\right]_D 
 \simeq m^2 \left( { \xi  \over M_4} \right)^2
 |\tilde{\Psi}|^2. 
\end{equation}
 while, for the gaugino mass, we have 
\begin{equation}
 \left[{X Y \over M_4^2} 
 {\mbox tr} \left(  {\cal W}^\alpha {\cal W}_\alpha \right) \right]_F 
 \simeq m \left( { \xi \over M_4} \right)^2  \lambda\lambda. 
 \label{gaugino mass}
\end{equation} 
Furthermore, the $\mu$-term can be obtained from 
\begin{equation}
 \left[{X^\dagger Y^\dagger \over M_4^2}  H_u H_d 
 \right]_{F^\dagger} \simeq 
 m \left( {\xi \over M_4} \right)^2 H_u H_d, 
 \label{mu-term}
\end{equation} 
 while the $B\mu$ term is obtained as 
\begin{equation}
\left[{X^\dagger X \over M_4^2} 
  H_u H_d \right]_D
 \simeq m^2 \left( {\xi \over M_4} \right)^2 H_u H_d.
 \label{Bmu}
\end{equation} 
Note that the relations  
 $\tilde{m}^2 \simeq B \mu$ and $M_{1/2} \simeq \mu$ 
 are automatically realized, 
 because of the $U(1)$ gauge invariance and 
 the holomorphy of the gauge kinetic function 
 and the superpotential. 
If we tune the parameters such that $\xi/M_4 = \delta \ll 1 $, 
 the split-SUSY mass spectrum, 
 $\tilde{m} \simeq m \delta \gg M_{1/2} \sim m \delta^2$, 
 is realized. 
In this case, the condition given in Eq.~(\ref{vanishing cc}) 
 implies the relations $m_{3/2} \simeq \tilde{m}/\sqrt{\epsilon} \gg \tilde{m}$. 
In summary, the above model leads to 
 a realistic split-SUSY mass spectrum satisfying 
 $m_{3/2} \gg \tilde{m} \simeq \sqrt{B \mu} \gg M_{1/2} \simeq \mu$ 
 under the condition 
 $ 0.01 \epsilon m_{3/2} \leq M_{1/2} = 100 \mbox{GeV}$--$1 \mbox{TeV}$, 
 for negligible anomaly mediation contributions.  
In the original paper, Ref.~\cite{Split-SUSY}, 
 the same split-SUSY mass spectrum is obtained 
 the basis of extra-dimensional models. 
In the following, we show that the model studied here 
 has more flexibility and 
 can lead to various split-SUSY mass spectra.

In general, we can introduce the usual (tree level) $\mu$-term 
 into the model as in the MSSM, that is, in the form 
\begin{eqnarray} 
\int d^2 \theta  \phi^3 \mu_{\mbox{tree}} H_u H_d .  
\end{eqnarray} 
Although this $\mu$ parameter can take any values,~\footnote{
 In the split-SUSY scenario, 
 it may not be so clear 
 whether the well-known $\mu$-problem is really a problem, 
 since this scenario is insensitive to fine-tunings. } 
 note that, once the $\mu$-term exists, 
 the relations 
 $B \mu \simeq F_\phi \mu_{\mbox{tree}} 
 \simeq \epsilon m_{3/2} \mu_{\mbox{tree}} $~\footnote{
 Here we have denoted the $\mu$ parameter at the tree level 
 as $\mu_{\mbox{tree}}$ to avoid confusion with the $\mu$-term 
 obtained in Eq.~(\ref{mu-term}).} is induced in SUGRA. 
Thus the total $\mu$ parameter ($\mu_{\mbox{total}}$) and 
 the total $B \mu$ ($(B \mu)_{\mbox{total}}$) are given by 
 $\mu_{\mbox{total}} \simeq M_{1/2} + \mu_{\mbox{tree}}$ and 
 $(B \mu)_{\mbox{total}} \simeq \tilde{m}^2+ \epsilon m_{3/2}\mu_{\mbox{tree}}
 \simeq \tilde{m}^2 +\sqrt{\epsilon} \tilde{m} \mu_{\mbox{tree}}$, 
 respectively. 
In the case $\mu_{\mbox{tree}} \leq M_{1/2}$, we obtain the above result. 
For the opposite case, $\mu_{\mbox{tree}} \geq M_{1/2}$, 
 the condition for a realistic split-SUSY scenario, 
 namely $\tilde{m}^2 + \mu_{\mbox{total}}^2 \sim (B \mu)_{\mbox{total}}$, 
 leads to $\mu_{\mbox{tree}} \leq \tilde{m}$ 
 and we obtain the mass spectrum $m_{3/2} \gg \tilde{m} \gg M_{1/2}$ 
 with $\mu_{\mbox{tree}}$ satisfying 
 $\tilde{m} \geq \mu_{\mbox{tree}} \geq M_{1/2}$. 

It is possible to extend our model to the case 
 in which the MSSM particles have non-zero $U(1)$ charges.  
We now show that this extended model can lead to 
 a split-SUSY mass spectrum that differs from that given above. 
Assume that $\xi \sim m$ with $g \sim 1$, for simplicity. 
In this case, the scalars in the MSSM  
 acquire mass through the VEV of D-term, 
 $\tilde{m}^2 = g q \langle D \rangle \sim q m^2$, 
 with the $U(1)$ charges. 
The condition given in Eq.~(\ref{vanishing cc}) implies 
\begin{eqnarray} 
 \tilde{m} \simeq m \simeq 
 \frac{\sqrt{\epsilon}}{\delta} m_{3/2}. 
\end{eqnarray} 
The gaugino mass is, again, obtained from Eq.~(\ref{gaugino mass}), 
 which yields $ M_{1/2} \sim m \delta^2$, and 
 thus we find $ M_{1/2} \sim \sqrt{\epsilon} \delta m_{3/2}$ 
 from the above equation. 
For $ \sqrt{\epsilon}/\delta \geq 1$ ($ \sqrt{\epsilon}/\delta \leq 1 $), 
 we obtain the mass spectrum 
 $ \tilde{m} \geq m_{3/2} \gg M_{1/2}$ 
 ($m_{3/2} \geq  \tilde{m} \gg M_{1/2}$). 

However, there is a problem in the above: 
 $B \mu$ obtained from Eq.~(\ref{Bmu}) is much smaller 
 than $\tilde{m}^2$, 
 and therefore the condition 
 for the realistic split-SUSY scenario cannot be satisfied. 
Unfortunately, the new contribution 
 $B \mu \simeq \epsilon m_{3/2} \mu_{\mbox{tree}}$ 
 from the tree level $\mu$ term cannot resolve this problem. 
This follows from the condition for a realistic split-SUSY scenario, 
 $\tilde{m}^2 + \mu_{\mbox{tree}}^2 
 \sim \epsilon m_{3/2} \mu_{\mbox{tree}}$. 
We cannot find any $\mu_{\mbox{tree}}$ satisfying this condition. 
A simple way to ameliorate the problem is to introduce 
 an additional contribution to $B \mu$. 
Let us consider a Polonyi model with the superpotential $W=M^2 \Phi$. 
We choose a special Kahler potential for $\Phi$ 
 so that the Polonyi model leads to 
 $\langle \Phi \rangle \simeq 0$ and $F_\Phi \simeq M^2$. 
In order not to change the structure of the $U(1)$ gauge sector, 
 $M^2$ must satisfy the condition 
 $M^2 \simeq F_{\Phi} \leq F_{X} \simeq m^2 $. 
Then, we introduce the superpotential for the Higgs sector 
 $ W_H = \Phi H_u H_d $, which leads to $B \mu \simeq M^2$. 
Then, tuning the parameter to realize 
 $M^2 \simeq \tilde{m}^2+ \mu_{\mbox{tree}}^2$, 
 we obtain realistic split-SUSY mass spectra in both cases, 
 $ \tilde{m} \geq m_{3/2}$ and $ m_{3/2} \geq \tilde{m}$, 
 with various values of $\mu_{\mbox{tree}}$ satisfying this condition. 

As discussed above, in the case with a tree level $\mu$-term, 
 we can take the $\mu$ parameter to be much larger 
 than the gaugino mass. 
This implies a mass spectrum different from 
 the originally proposed in Ref.~\cite{Split-SUSY}. 
The phenomenology of the split-SUSY scenario 
 with such a large $\mu$-parameter 
 is investigated in Ref.~\cite{SSplit}.  
However, we note that, once a large $\mu$ term is introduced, 
 the Higgs superfields play the role of the ``messengers'' 
 in the gauge mediated SUSY breaking model 
 (gauge mediation)~\cite{GMSB}, 
 and as a result, the gauginos (wino and bino) acquire soft masses 
 of the order of $(\alpha_{SM}/4 \pi) B \mu / \mu$. 
Hence, the scale of $\mu$ is limited in order to keep 
 the gaugino masses near the electroweak scale. \\

{\bf 2. Case of small gravitino mass ($0.01 \; m_{3/2} < M_{1/2}$)}\\

\noindent
In this case,  
 the contribution from the anomaly mediation is small, 
 and therefore the almost no-scale structure is no longer necessary. 
The split-SUSY mass spectrum in this case implies 
 $\tilde{m} \gg m_{3/2}$. 
Therefore, a SUSY breaking mediation mechanism 
 other than the SUGRA mediation 
 should have the dominant contribution to sparticle masses. 
Again, let us consider the $U(1)$ gauge model, 
 in which the MSSM matter and Higgs superfields 
 have non-zero $U(1)$ charges. 
As discussed above, large values of the soft mass squared 
 for the scalars in the MSSM are induced through the $U(1)$ D-term. 
The main difference between this and the previous model 
 is that here, the almost no-scale structure is no longer necessary. 
The soft mass spectrum in this case is obtained 
 by taking $\epsilon \sim 1$ in the previous results, 
 Eqs.~(\ref{vanishing cc}) and (\ref{gaugino mass}); 
 this yields $\tilde{m} \simeq m \gg m_{3/2} \simeq m \delta \gg 
 M_{1/2} \sim m \delta^2$. 
New contributions to the gaugino masses are necessary 
 to realize the case $M_{1/2} > m_{3/2}$. 
In order to obtain these contributions, 
 let us introduce the gauge mediation sector 
 into the model~\cite{GMSB}. 
Consider a simple messenger sector given by 
\begin{equation}
\label{W_m}
 W_m  = \left( M_m+ F_m \theta^2 \right)  
 \left( \lambda_Q \overline{Q} Q  +
        \lambda_\ell \overline{\ell} \ell
 \right),  
\end{equation}
where $\overline{Q}$, $Q$, $\overline{\ell}$ and $\ell$ 
 are the vector-like messenger quarks and leptons, respectively. 
The MSSM gaugino masses are generated through one-loop diagram 
 of the messenger fields as 
\begin{equation}
 M_{1/2} \simeq \left(\frac{\alpha_{SM}}{4 \pi} \right)  
 {F_m \over M_m}.
\label{GMSB-gaugino}
\end{equation}
If $\langle F_X \rangle \geq F_m$, 
 the gravitino mass retains the same value, 
 while the gaugino mass can become larger than the gravitino mass 
 through the gauge mediation 
 when the messenger scale, $M_m$, is small enough. 
Choosing appropriate values for $M_m$ and $F_m$ 
 in the messenger sector, 
 we can realize the split-SUSY mass spectrum 
 $\tilde{m} \gg M_{1/2} \gg m_{3/2}$ 
 with a small gravitino mass. 

For the Higgs mass parameters, 
 we can use the chiral superfield $M_m + F_m \theta^2$  
 in the messenger sector and introduce the Higgs superpotential 
\begin{equation}
 W_H = \left( M_m + F_m \theta^2 \right) H_u H_d,  
\end{equation} 
 in addition to the $\mu$-term at the tree level.   
Thus, we obtain $\mu_{\mbox{total}} = M_m + \mu_{\mbox{tree}}$ 
 and $ B \mu \simeq F_m$. 
A realistic split-SUSY mass spectrum requires 
 $ \tilde{m}^2 + (M_m+\mu_{\mbox{tree}})^2 
 \simeq B \mu \simeq F_m$, 
 which can be rewritten as 
  $ \tilde{m}^2 + (M_m+\mu_{\mbox{tree}})^2 \simeq 
  100 M_{1/2} M_m $ 
 by using Eq.~(\ref{GMSB-gaugino}). 
In order to realize the hierarchy $\tilde{m} \gg 100 M_{1/2}$ 
 ($M_m \gg 100 M_{1/2}$), 
 a cancellation between $M_m$ and $\mu_{\mbox{tree}}$ is necessary 
 so as to satisfy $ (M_m+\mu_{\mbox{tree}})^2 \leq 100 M_{1/2} M_m$.
Once this fine-tuning is realized, 
 we can obtain a realistic split-SUSY mass spectrum 
 with $\tilde{m} \simeq B \mu \gg M_{1/2}$, 
 with the $\mu$ parameter satisfying 
 $\tilde{m} \geq \mu \geq M_{1/2}$

\section{Summary}

In a recently proposed split-SUSY scenario~\cite{Split-SUSY},
 the scalar masses are very large, 
 while the fermion masses, protected by chiral symmetry, 
 are set to the electroweak scale. 
In this paper, we have shown, in detail, 
 how the Higgs sector of the MSSM is reduced to that of the SM 
 in the limit of large soft SUSY breaking 
 Higgs mass parameters.  
Then, we demonstrated that the conditions  
 $|{m}_{u}^2|\sim |{m}_{d}^2|\sim |B\mu |$ 
 are necessary to obtain a suitable magnitude of $\tan \beta$ 
 that yields a realistic bottom quark Yukawa coupling. 
Based on these conditions, we have presented some simple models 
 that provide realistic split-SUSY mass spectra
 for various gravitino mass scales, 
 from $m_{3/2} \gg \tilde{m}$ to $M_{1/2} \gg m_{3/2}$.

\vskip 1cm

\leftline{\bf Acknowledgments}

This work is supported in part by Scientific Grants from 
 the Ministry of Education and Science in Japan
 (Grant Nos.\ 14740164, 15740164, 16028214, and 16540258). 
N.O. would like to thank the Theoretical Elementary 
 Particle Physics Group of the University of Arizona, 
 and especially, Zackaria Chacko and Hock-Seng Goh, 
 for their very kind hospitality 
 during the completion of this work. 

\vspace{.5cm}

\appendix
\section{Mass spectra in split-SUSY (MSSM)}

In this appendix, we show 
 that $\widetilde{h}$ can be 
 regarded as the SM Higgs doublet in the
 split-SUSY scenario  
 by determining the masses of 
 the charged and neutral sectors. 
The Higgs doublets are written 
\begin{equation}
H_u=
\left(
\begin{array}{c}
H_u^+ \\
{1\over\sqrt{2}} (v\sin\beta + \eta_u + i \zeta_u)
\end{array}
\right), \;\;\;
H_d=
\left(
\begin{array}{c}
{1\over\sqrt{2}} (v\cos\beta + \eta_d + i \zeta_d) \\
H_d^-
\end{array}
\right), 
\label{HuHd}
\end{equation}
and we carry out a field redefinition 
through which we introduce 
 ${H_1}^T\equiv(\epsilon H_d^*)^T =
 \left(
  H_d^+ , 
  -{1\over\sqrt{2}} (v\cos\beta + \eta_d - i \zeta_d) 
 \right)$ 
and $H_2\equiv H_u$, as in section 2. 
Equation (\ref{HuHd}) suggests that 
 the direction of the VEV 
 is the same as that of $\widetilde{h}$, defined in 
 Eq.~(\ref{H}). 
Therefore, the light field $\widetilde{h}$ is just 
 a linear combination of the fields acquiring the VEV. 
The direction of the would-be NG bosons 
 ($W^\pm, Z$) is
 also the same as that of $\widetilde{h}$. 
This can be understood
 by considering the following charged and pseudo-scalar
 mass matrices:  
\begin{eqnarray}
-{\cal L}_{\mbox{charged}} &=& 
(\tilde{m}^2+M_W^2)(H_d^-, H_u^-)
\left(
\begin{array}{cc}
\sin^2\beta & \sin\beta\cos\beta \\
\sin\beta\cos\beta & \cos^2 \beta
\end{array}
\right) 
\left(
\begin{array}{c}
H_d^+ \\
H_u^+
\end{array}
\right), \\
-{\cal L}_{\mbox{pseudo}} &=& 
{\tilde{m}^2\over2}(\zeta_d, \zeta_u)
\left(
\begin{array}{cc}
\sin^2\beta & \sin\beta\cos\beta \\
\sin\beta\cos\beta & \cos^2 \beta
\end{array}
\right)
\left(
\begin{array}{c}
\zeta_d \\
\zeta_u
\end{array}
\right).
\end{eqnarray}
These matrices suggest that the 
 physical charged (pseudo-scalar) Higgs is 
 in the direction of $\widetilde{H}$, and its mass is
 ${m}^2_A+M_W^2$ (${m}^2_A$). 
This mass is ${\mathcal O}(\tilde{m})$,
 which is too large to survive
 at low energy in the split-SUSY scenario.
On the other hand, 
 the mass matrix of the neutral scalar 
 is given by 
\begin{eqnarray}
\label{12a}
-{\cal L}_{\mbox{scalar}} &=& 
{1\over2}(-\eta_d, \eta_u)
\left(
\begin{array}{cc}
{m}_A^2\sin^2\beta +M_Z^2\cos^2\beta
   & ({m}_A^2+M_Z^2)\sin\beta\cos\beta \\
({m}_A^2+M_Z^2)\sin\beta\cos\beta 
   & {m}_A^2\cos^2\beta +M_Z^2\sin^2\beta
\end{array}
\right) 
\left(
\begin{array}{c}
-\eta_d \\
\eta_u
\end{array}
\right), \\
 &=& 
{{m}_A^2\over2}(-\eta_d, \eta_u)
\left(
\begin{array}{cc}
\sin^2\beta &  \sin\beta\cos\beta \\
\sin\beta\cos\beta & \cos^2 \beta
\end{array}
\right) 
\left(
\begin{array}{c}
-\eta_d \\
\eta_u
\end{array}
\right) + {\mathcal O}(M_Z^2) .
\end{eqnarray}
This implies that the mass matrix 
 can be diagonalized 
 by the same 
 linear combination 
 {\it to a good approximation} 
 in the split-SUSY scenario.
In the case $M_Z^2 \ll {m}_A^2$,  
 the light and heavy neutral Higgs eigenstates, 
 $h^0$ and $H^0$,
 are given by $h^0=-\cos\beta (-\eta_d) + \sin\beta \eta_u$  
 and $H^0=\sin\beta (-\eta_d) + \cos\beta \eta_u$, respectively, 
 which means that the light neutral scalar $h^0$ is included 
 in $\widetilde{h}$.
Without the approximation, 
 Eq.~(\ref{12a}) becomes 
\begin{eqnarray}
\left(
\begin{array}{cc}
M_Z^2 \cos^2(2\beta) & {M_Z^2\over2}\sin(4\beta) \\
{M_Z^2\over2}\sin(4\beta) & 
 m_A^2 + {M_Z^2\over2}(1-\cos(4\beta)) 
\end{array}
\right) 
\label{exact}
\end{eqnarray}
in the basis $({h^0}, {H^0})$.
Because the off-diagonal elements are ${\mathcal O}(M_Z^2)$, 
 the heavy field ${H^0}$ with mass ${m}_A$
 decouples, and only 
 ${h^0}$ survives at low energy, 
 having mass $M_Z^2 \cos^2(2\beta)$
 when $M_Z^2 \ll m_A^2$. 

In summary, 
 we have shown that $\widetilde{h}$ 
 is in the direction as the VEV and 
 containing would-be NG boson and also a light scalar. 
Therefore, we can conclude that 
 $\widetilde{h}$ corresponds to the SM Higgs doublet,  
 which is what we sought to show in this appendix. 
In the split-SUSY scenario,
 $\widetilde{H}$ is decoupled,  and
 only $\widetilde{h}$ survives at low energy.

\vspace{1cm}

%
%

%
\end{document}